# CHARGE POLARIZATION EFFECTS AND *HOLE* SPECTRA CHARACTERISTICS IN $Al_xGa_{1-x}N/GaN$ SUPERLATTICES

Fatna Assaoui
*Department of Physics University Mohammed V, Scientific Faculty,*
*Av. Ibn Battouta, B.P. 1014 Rabat, Morocco*
*and*
*The Abdus Salam International Centre for Theoretical Physics, Trieste, Italy*

and

Pedro Pereyra[1]
*Depto. de Ciencias Básicas, UAM-Azcapotzalco,*
*Av. S. Pablo 180, C.P. 02200, México D.F., México*
*and*
*The Abdus Salam International Centre for Theoretical Physics, Trieste, Italy.*

## Abstract

We study the effects of charge polarization on the extended physical properties of superlattices, such as transmission coefficients and valence band structure. We consider both linear and parabolic modulation of the band edge. Based on the theory of finite periodic systems (TFPS), analytic expressions and high precision calculations of the relevant physical quantities for $n$-cell systems are obtained. New and also well-known features of these systems are identified. Besides the well-known energy bandstructure, we also have the *field bandstructure*, with interesting characteristics. Wider field gaps at stronger internal electric fields and higher density of field bands for larger layer widths are some of these characteristics. Well defined level density asymmetries identify the minibands induced by charge polarization or the so-called Quantum Confining Stark Effect. We present the $n$-cell transmission amplitudes, transmission coefficients and miniband structures for different values of the relevant parameters.

[1] Regular Associate of the Abdus Salam ICTP.

## I. INTRODUCTION

The spontaneous dielectric polarization and the piezoelectric response observed in $Mg$ doped $(Al_xGa_{1-x}N/GaN)^n$ and $(In_xGa_{1-x}N/GaN)^n$ superlattices, and heterostructures, lead to the existence of localized 2D electron and hole gases on the opposite interfaces of the quantum wells[1]. Therefore, to the conduction- and valence-band bending. This effect denoted as the quantum confined Stark effect (QCSE) has important consequences on the extended superlattice properties like the miniband structure, intraband eigenfunctions and eigenvalues, and the intra- and inter-band transitions[1–14]. The understanding and description of this effect is important for the overwhelming number of applications, both in optoelectronic and electronic, based on the nitrides' properties and their emission spectra.

To study the effects of charge polarization on the transmission properties and band structure, we shall consider two types of potential profiles: one with linear and the other with parabolic modulation of the band-edge. Solving the single-cell problem, and using the rigorous and compact formulas of the theory of finite periodic systems (TFPS)[15], we can obtain analytic expressions and perform high precision calculations of scattering amplitudes and the resonant band structure. We will present the $n$-cell transmission amplitudes, transmission coefficients and miniband structures from different points of view: As a function of the energy and also as a function of the internal electric field strength. We analyze the effect of the layer width, especially on the reduction of the recombination energy. We will also show that, in the linear case and for a fixed Fermi energy, a very appealing field bandstructure is obtained when the internal electric field is varied. Our purpose is to offer a theoretical description of the way in which the extended properties depend on the internal electric field and on the layer widths.

For these systems, three characteristic energy regions can be distinguished. In each region the consequences on the extended physical properties are clearly recognized. In the lowest energy region ($E < aF$ in the linear case and $E < E_1$ in the parabolic case, see figure 1), the potential barrier effectively increases and pushes up the allowed states and reduce their density. Depending on the specific physical parameters, some extremely thin and stable minibands (with band-widths of the order of $10^{-7}eV$) are found. In the highest energy region the potential barrier effectively diminishes, and as a consequence more and wider minibands appear.

We shall present here the principal theoretical expressions that will then be applied to our specific and particular cases. In Section II we obtain some general results for the transmission amplitudes for the linear and parablic potential modulations. In section III we present the transmission coefficients and the minibands as a function of the electric field and other superlattice parameters and conclude with a discussion on the results.

## II. TRANSMISSION COEFFICIENTS AND BANDSTRUCTURE

To calculate the transmission amplitudes we need to solve the Schrödinger equation of the multilayer system, in the one channel 1D approximation commonly used when the transverse translational invariance holds. We are interested in solving that equation for the specific superlattice potential profiles shown in figure 2, where the well-known square barrier potential shapes (produced by the alternating semiconductor layers) are modified by the internal electric field generated by charge polarization. Non-linear modulation of the potential profile has also been suggested recently by Goepfert et al. after solving numerically the Poisson's equation[3]. In the linear case, and in the effective mass approximation, the valence band potential will be taken as (the subindices $l$ and $h$ stand for low and high potential regions)

$$V_l(z) = zF \qquad \text{for} \quad z < a$$

and

$$V_h(z) = -zF + \Delta E_c + 2aF \qquad \text{for} \quad z > a,$$

In the non-linear case the potential will be taken as

$$V_l(z) = -\frac{E_1}{z_o^2}(z - z_o)^2 + E_1 \qquad \text{for} \quad z < a$$

and

$$V_h(z) = 4\frac{\Delta E_v - E_1}{a^2}(z - z_o - a)^2 + E_2 \qquad \text{for} \quad z > a.$$

Here $\Delta E_v$ refers to the valence band offset, $F = e\mathcal{E}$ is the electric force, $a$ the layer width, and $z_o = a/2$. The parameters $E_1, E_2$ and $F$ depend on the charge polarization strength. To solve the Schrödinger problem and to understand the quantum confined Stark effect, we use the strategy and formulas of the theory of finite periodic systems[16], i.e. we first solve the single-cell problem and obtain the single-cell transmission amplitude $t$, we then use relations like

$$t_n = \frac{t^*}{U_{n-1}t^* - U_{n-2}}$$

to determine physical quantities of the whole system, i.e. to determine the extended physical properties, here $U_n$ is the Chebyshev polynomial of the second kind. The superlattice bandstructure maintains a close relation with the resonant behavior of the superlattice transmission coefficient

$$T_n = \left| \frac{t^*}{U_{n-1}t^* - U_{n-2}} \right|^2$$

This quantity provides all the information on the allowed and forbidden energy regions as well as on the internal electric fields where the particle is transmited or becomes localized. It is then of great and relevant value to study the way in which the different parameters affect the $n$-cell trasmission coefficient. Using the previous analytic formulas, these quantities, including the intraband energy levels, can be calculated accurately.

All we need is to solve the single-cell Schrödinger problem as exactly as possible. It is well known that using the WKB aproximation, the wave functions are written either as

$$\varphi(z) = a\ e^{i\int^z p(z)dz} + b\ e^{-i\int^z p(z)dz} \tag{1}$$

with $p(z) = \sqrt{2m(E - V(z))/\hbar^2}$ for the classically allowed energy regions or as

$$\varphi(z) = a\ e^{\int^z q(z)dz} + b\ e^{-\int^z q(z)dz} \tag{2}$$

with $q(z) = \sqrt{2m(V(z) - E)/\hbar^2}$ for the classically forbidden regions.

For both the linearly modulated potential in figure 1(a) and the non-linear potential in figure 1(b), it has been possible to obtain the analytic expression for the transmission amplitudes at each of the energy regions. To illustrate and introduce the notation we write here the transmission amplitude for the lowest energy region of the potential shown in figure 1(a), which is given by

$$\begin{aligned} t = e^{-iP_{0a}}&[e^{Q_{l,z_1z_2}+Q_{h,z_1z_2}}(1+i)(r_{lh,a}+q_{ha}+q_{la})(-r_{hl,2a}+q_{h2a}+ik_{h2a}) \\ &-e^{-Q_{l,z_1z_2}-Q_{h,z_1z_2}}(1-i)(-r_{lh,a}+q_{ha}+q_{la})(-r_{hl,2a}+q_{h2a}-ip_{h2a}) \\ &e^{-Q_{l,z_1z_2}+Q_{h,z_1z_2}}(1-i)(r_{lh,a}+q_{ha}-q_{la})(-r_{hl,2a}+q_{h2a}+ip_{h2a}) \\ &-e^{Q_{l,z_1z_2}-Q_{h,z_1z_2}}(1+i)(r_{lh,a}+q_{ha}-q_{la})(-r_{hl,2a}+q_{h2a}-ip_{h2a})]^{-1} \\ &(4i\sqrt{p_{l2a}\ q_{h2a}\ q_{ha}\ p_{la}}) \end{aligned} \tag{3}$$

Here

$$P_{j,z_1z_2} = \int_{z_1}^{z_2} p_j(z)dz,$$

$$Q_{j,z_1z_2} = \int_{z_1}^{z_2} q_j(z)dz,$$

$$p_{jz} = p_j(z) = \sqrt{2m(E - V_j(z))/\hbar^2} \qquad j = l, h$$

$$q_{jz} = q_j(z) = \sqrt{2m(V_j(z) - E)/\hbar^2} \qquad j = l,$$

and

$$r_{lh,a} = \frac{1}{2\hbar^2}\left(\frac{m_l^* V_l'(a)}{p_l^2(a)} + \frac{m_h^* V_h'(a)}{q_h^2(a)}\right)$$

Similar expressions are obtained for each energy region and for the two types of potential profiles. Given these transmission amplitudes we are able to determine the Superlattice band-structure and the $n$-cell tranmission coefficients.

Before discussing the $n$-cell results it is worth having a look at the behavior of the single-cell transmission coefficient. In figure 3 we have the single-cell transmission coefficients for both types of potential profiles. The charge polarization effect is visible precisely in the energy region where the band-edge modulation (BEM) occurs. For the linear case, the reduction of the potential barrier leads to the appearance of a giant resonance, which for superlattices will become a miniband. In the single-cell problem the monotonous increasing of the transmission amplitude breaks (see figure 3(a)). A similar resonance appears for the non-linear case. For this system, the single-cell transmission coefficient has a resonant behavior due to the shallow potential well in the upper part of the potential barrier. This effect is certainly related with the high-performance and conductance enhancement experimentally observed[7,6,10,9,8].

## III. INTERNAL ELECTRIC FIELD EFFECTS ON THE $N$-CELL PROPERTIES

In this Section, we will present some results for the bandstructure and transmission coefficients for $n$-cell $Al_xGa_{1-x}N/GaN$superlattices. We shall start (figures 4, 5 and 6) analyzing the bandstructure, similarities and differences, in the hole-spectra for the two types of band-edge modulation considered here (linear and parabolic) and with the reported results. We then look, for fixed layer width, at the field effect on the band structure (figure 7) and present the very appealing resonant behavior of the transmission coefficient as function of the electric field. Simple but illustrative examples of the field-bandstructure will be presented in figures 8 and 9. We discuss the reduction of the emission energy and finally we shall present the transmission coefficients for the linear and parabolic potential profiles for differente layer widths and different internal electric field intensity..

In figures 4 and 5, we present the bandstructure for both linear and non-linear modulation of the band-edge. In figure 4, we consider a potential profile with parabolic modulation and in figure 5 the linear case. For these figures we choose the parameters in such a way that we can compare between the two types of potential and also with the specific examples reported in the literature for $Al_{0.2}Ga_{0.8}N/GaN$ superlattices. The layer width in both cases is $a = 20$Å. While in the linear case we shall consider a valence band offset $\Delta E_v = 0.12eV$, (which corresponds to $x = 0.2$ when the bowing parameter is 1.3). In this case the highest point in the potential barrier is at $\Delta E_v + aF$. For $a = 20$Å and $F = 0.001$ this maxima is at $E_b = 0.14eV$. In the parabolic case we will consider the parameters used in Ref. [3] with maxima and minima of the parabolas at $E_1 = 0.012eV$ and $E_2 = 0.127eV$, respectively and the barrier height at $E_b = 0.14eV$. These systems are to some extent equivalent. What do we obtain for their bandstructures?

In the parabolic case, for energies between 0 and $0.15eV$ we have two minibands while in the linear case we have three. In both cases there is a miniband in the energy region of the upper part of the potential barrier, where the barrier width diminishes. The presence of these minibands depends entirely on the modulation of the band-edges. Very precise experiments may identify these minibands because there is an interesting feature in this type of minibands. The density of levels is asymmetric with higher density in the upper miniband edge. While in the minibands of the intermediate region the level density is basically symmetric. In figure 6, we plot the transmission coefficients in a normal miniband and the transmission coefficient in the miniband induced by the band edge modulation. The bands in the upper parts of the barrier and those in the intermediate energy regions of the parabolic and linear cases in figures 4 and 5 ($E_1 < E < \Delta E_v$,and $aF < E < \Delta E_v + aF$, respectively) have similar position and widths. The most important difference between these two figures is the presence in the linear case and the absence in the parabolic, of a very thin low energy band of highly stable states. In figure 7, this miniband is pushed up when the electric field is increased.

We notice that the position of the lowest band in Ref. [2] corresponds aproximately to our second band in figure 5. In figures 4 and 5 the bandstructures were plotted for finite superlattices. In this case, for $n = 14$. It is well known that increasing the number of cells $n$ the only effect that results is the increasing of the intraband level density. It is worth noticing that the level density in the minibands, in the upper part of the potential region, is asymmetric, with higher density at higher energies.

As shown in figure 7, interesting band displacements are observed when the electric field strength varies. Increasing the electric field, the minibands in the intermediate energy region ($aF < E < aF + \Delta E_v$) shift to higher energies while the band widths remain almost constant. Changing $F$ from $0.0001 eV/Å$ to $0.001 eV/Å$, the bands experience a displacement of $20 meV$. In the high energy region ($E_2 < E < \Delta E_v$) we observe that, while the shift is still toward higher energies, their width grows appreciably.

Fixing the particle's energy, other interesting property in the field dependent bandstructure is found. This property, reminiscent of the optical bandstructure, is related to the bandstructure dispacement produced when the internal electric field $\mathcal{E} = F/e$ or, equivalently, when the charge concentration changes. In figures 8 and 9, transmission coefficients for different energies and layer widths are plotted as functions of the electric field. In these figures it is apparent a bandstructure displacement when the particle's energy varies. In figure 8 the particle's energy is kept constant. Big changes in the *field-band-structure* can be observed. As could be expected, for wider layers the bands are thinner and move to lower energies. The bands at higher internal electric fields are also thinner than those at lower electric fields, eventually the transmission coefficient vanishes and the particle localizes with no transmission. On the other hand, if we now fix the layer width, as in figure 9, and vary the particle's energy, we observe that by increasing linearly the energy, the *field-band-structure* displaces alsmost linearly with rather small effects in the band widths. Notice that for electric fields beyond, say $E/a$, there is no more transmission, and again the particle localizes.

An interesting feature of the QCSE mentioned in the literature is the strong dependence of the emission energy on the quantum well thickness[4,5]. This effect can be observed (see figure 10) when the electric field is fixed and we determine the bandstructure for different layer widths. The emission energy depends on the electron and hole spectra as well as on the energy distance between the valence and the conductance bands, which as seen in figure 11 become closer in the presence of internal electric fields. In principle, the change in the emission energy is basically given by

$$-aF + \Delta E_e + \Delta E_h$$

For $F \approx 10^{-4}$ and $a = 20\text{Å}$, the contribution of the first term is of the order of $2meV$ while the *hole* band shift $\Delta E_h$ is of the order of $10meV$. This agrees well with the experimental results in Ref.[5]. Notice and recall that when the *hole*-miniband moves to lower energies the electron miniband moves to higher energies, reducing the emission energy.

Finally, in figure 12 we have the transmission coefficient plotted as functions of the particles energy for different internal electric fields, i.e. different polarized charge concentrations. In this figure we fix the layer width as $a = 100\text{Å}$. Increasing the internal electric field the product $aF$ grows and the potential profile modifies as can be seen in $12(a)$. As a consequence the energy levels are pushed up and, at the same time, the minibands in the upper part of the potential barrier become narrower. To observe minibands at lower energies we need to reduce the barrier width as can be seen in figure 5 where we show the transmission coefficients for $a = 20\text{Å}$.

In general, both types of potential profiles, the linear and the parabolic potential modulation, with almost equivalent physical parameters in the sense mentioned above, provide trasmission coefficients and bandstructures with similar features, except at very low energies.

## IV. CONCLUSIONS

We solved the superlattice Schrödinger equation for systems whose potential profile is modified by charge polarization in opposite interfaces. The band structure and transmission coefficients are calculated with the highest precision possible. While in the lower and intermediate energy regions narrow and thin minibands occur, in the upper energy regions the number of minibands increases and their widths also. With the solutions reported here, one can determine band displacements, the intra-band energy levels (including the resonance widths) for different layer widths and internal electric fields, hence the time of life of the excited particles. The results obtained here are relevant for design of optoelectronic and electronic-devices.

## V. ACKNOWLEDGMENTS

We acknowledge the Abdus Salam International Centre for Theoretical Physics, Trieste, Italy, for hospitality and financial support during our visits where this work was performed. One of us (P. P.) acknowledges the partial support of CONACyT México project No. E-29026. This work was done within the framework of the Associateship Scheme of the Abdus Salam ICTP.

## REFERENCES

[1] P. Kozodoy, Monica. Hansen, Steven. P. Denbaars and Unesh K. Mishra, Appl. Phys. Lett. **74**, 3681 (1999).

[2] S. Hackenbuchner, J. A. Majewski, G. Zandler, G. Vogl, J. Crystal. Growth. **230**, 607 (2001).

[3] D. Goepfert, E. F. Schuber, A. Osinsky, P. E. Norris and N. N .Faleev, Appl. Phys. Lett. **88**, 2030 (2000).

[4] P. Perlin, S. P. Lepkowski, H. Teisseyre, T. Suski N. Grandjean and J. Massies, Acta. Physica. Polinica A. **100**, 261 (2001).

[5] S.P. Lepowsky, H. Teisseyre, T. Suski, P. Perlin, N. Grandjean and J. Massies, *Appl. Phys. Lett.* in print

[6] G. Parish, S. Keller, P. Kozodoy, J. P. Ibbetson, H. Marchand, P. T. Fini, S. B. Fleisher, S. P. DenBaars, U. K. Mishra and E. J. Tarsa, Appll. Phys. Lett. **75,** 247 (1999).

[7] P. Kozodoy, Yulio P. Smorchkovo, Monica. Hansen, Huili Xing, Steven. P. Denbaars, Unesh K. Mishra, A. W. Sascler, R. Perrin and W. C. Mitchel, Appl. Phys. Lett. **75**, 2444 (1999).

[8] N. Grandjean, J. Massiers, S. Dalmasso, P. Vennegues, L. Siozade and L. Hirsch, Appl. Phys. Lett. **74**, 3616 (1999).

[9] R. Langer, A. Barski, J. Simon, N. T. Pelekanos, O. Konovalov, R. Andre and Le Si Dang, Appl. Phys. Lett. **74**, 3610 (1999).

[10] C. H. Kuo, J. K. Sheu, G. C. Chi, Y. C. Huang and T. W. Yeh, Solid. state. Elctronic. **45**, 717 (2001).

[11] H. X. Jiang, S. X. Jin, J Li, J. Shakya and J. Y. Lin, Appl. Phys. Lett. **78**, 1303 (2001).

[12] S. Einfeldt, H. Heinke, V. Kirchner and D. Hommel, Appl. Phys. Lett. **89**, 2160 (2001).

[13] Aldo Dicarlo, Fabio Della Sala, Paolo Lngli, Vincenzo Fiorentini and Fabio Bermardini, Appl. Phys. Lett. **76**, 3950 (2000).

[14] Lee. Cheul-Ro, Son. Sung-Jin, Seol. Kyeong-Won, Yeon. Jeong-Mo, Haeng- Keun and Park. Yong-Jo, J. Crystal. Growth. **226**, 215 (2001), Lee. Cheul-Ro, J. Crystal. Growth. **220**, 62 (2001).

[15] P. Pereyra, Phys. Rev. Lett. **804** 2677 (1998).

[16] P. Pereyra, j. Phys.. AÑ Math. Gen. **31**, 4521 (1998), P. Pereyra, E. Castillo, Submitted Phys. Rev. B, Preprint IC/2001/122, ICTP, Trieste, Italy.

## FIGURES

FIG. 1. The linear and parabolic modulation of the potential profile produce additional repulsion and a wider potential barrier in the lower energy part and a smaller barrier width in the upper parts.

FIG. 2. Superlattices with linear (a) and parabolic (b) band-edge modulation studied in this work.

FIG. 3. The reduction of the barrier width in the upper part of the barrier leads to a beautiful resonant effect in single-barrier transmission coefficients. In (a) the potential is linear. In (b) we have the transmission for a potential barrier with parabolic modulation. Due to the valley produced in its upper part, the single cell transmission coefficients have a resonant behavior.

FIG. 4. The band structure in the parabolic modulated superlattice. The potential parameters are indicated in the figure. In (c) we have a characteristic bandstructure in the upper part of the potential. The asymmetry, reflected in the increase of resonant states at higher energies, is a consequence of the barrier narrowing.

FIG. 5. The band structure in the linear modulated superlattice. The potential parameters are indicated in the figure. In this case a very narrow band of quite stable states, shown in (c), is present.

FIG. 6. A symmetric and an asymmetric band. The asymmetric bands, at energies close to the band offset, are characteristic of these types of systems.

FIG. 7. The blue shift of the *hole*-spectra when the internal electric field increases.

FIG. 8. A band structure as a function of the internal electric field for different values of the layer width $a$. In these graphs the particle energy is kept fixed.

FIG. 9. A band structure as a function of the internal electric field for different values of the particle's energy. In these graphs the layer width is kept fixed.

FIG. 10. The reduction of the recombination energy in the charge-polarized superlattices observed here when the layer width increases. At the same time the band-widths become extremely narrow and the associated states more stable.

FIG. 11. The conduction and the valence-band edges modulation and its effect on the recombination energy. A desplacement to higher energies in the *hole*-spectra implies a displacement of the electron states in the valence band, and viceversa.

FIG. 12. The bandstructures for different potential profiles induced by different degrees of charge polarization strengths.

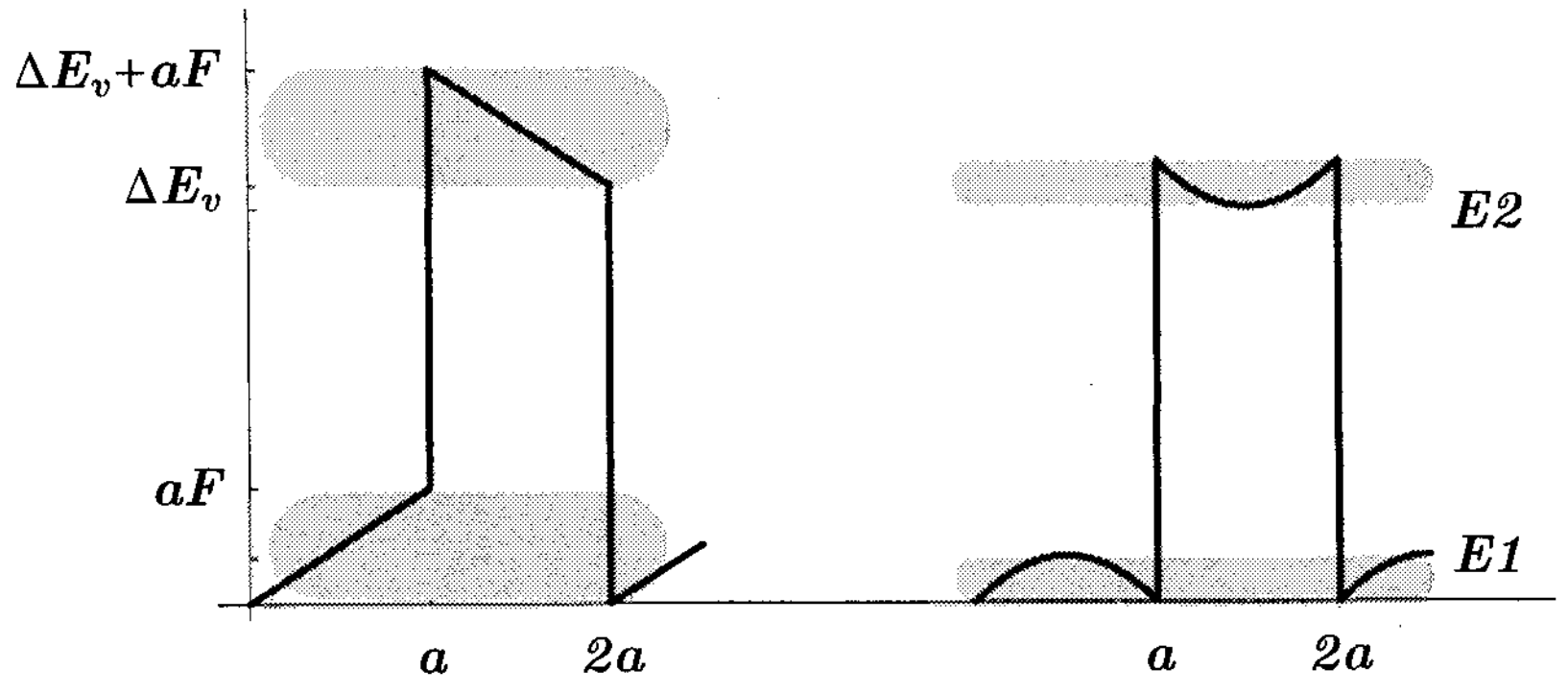

Fig.1

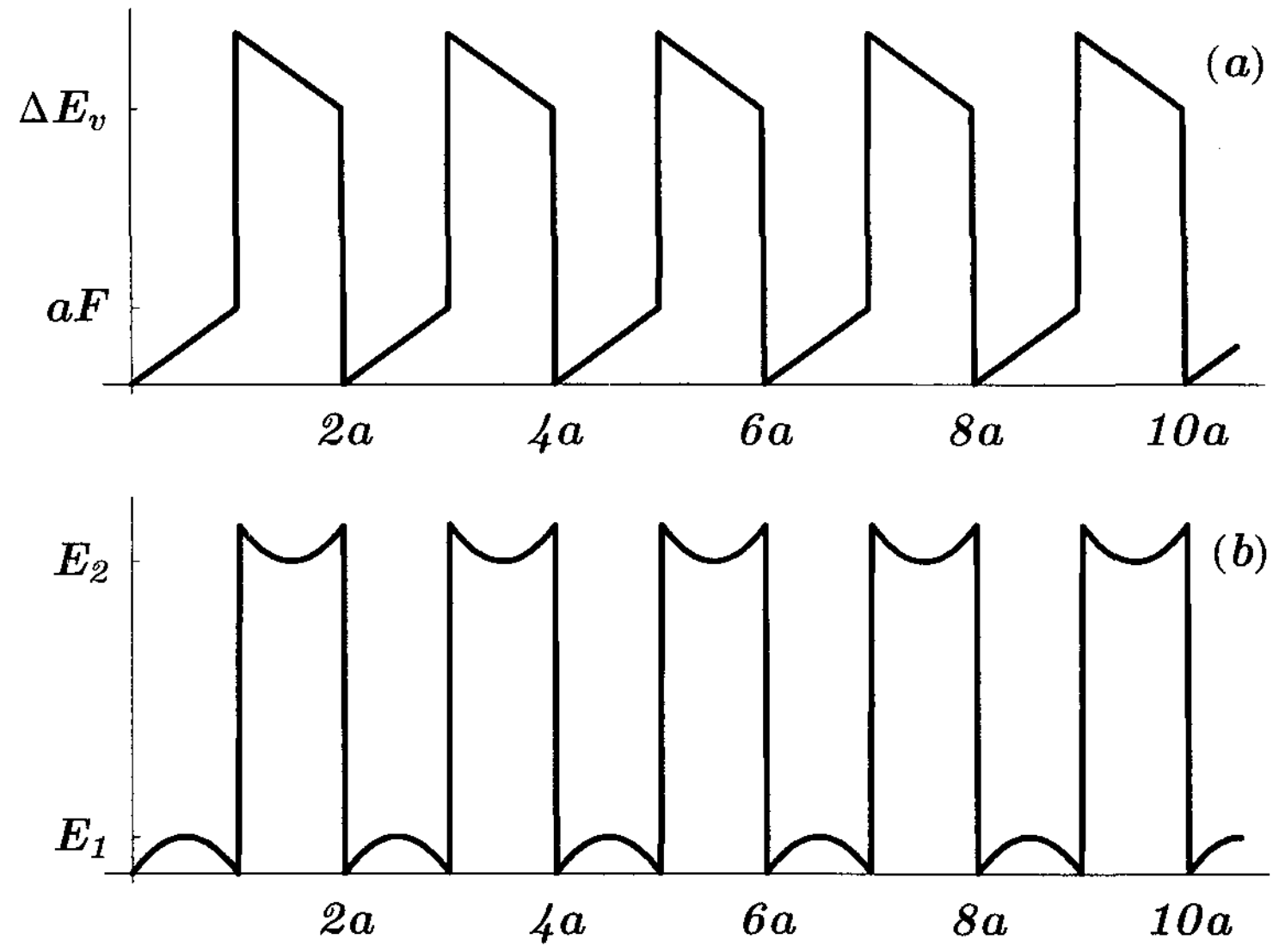

Fig.2

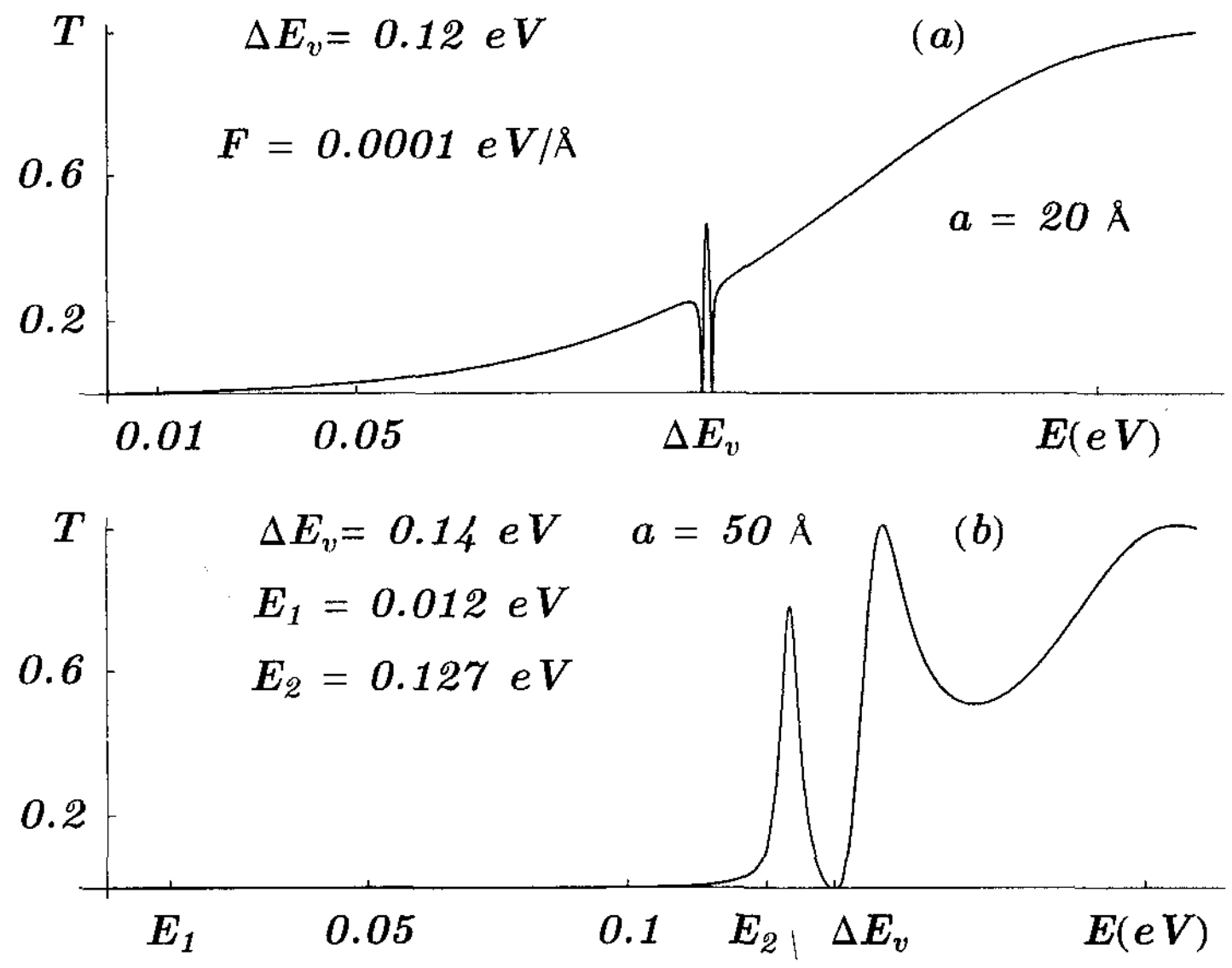

Fig.3

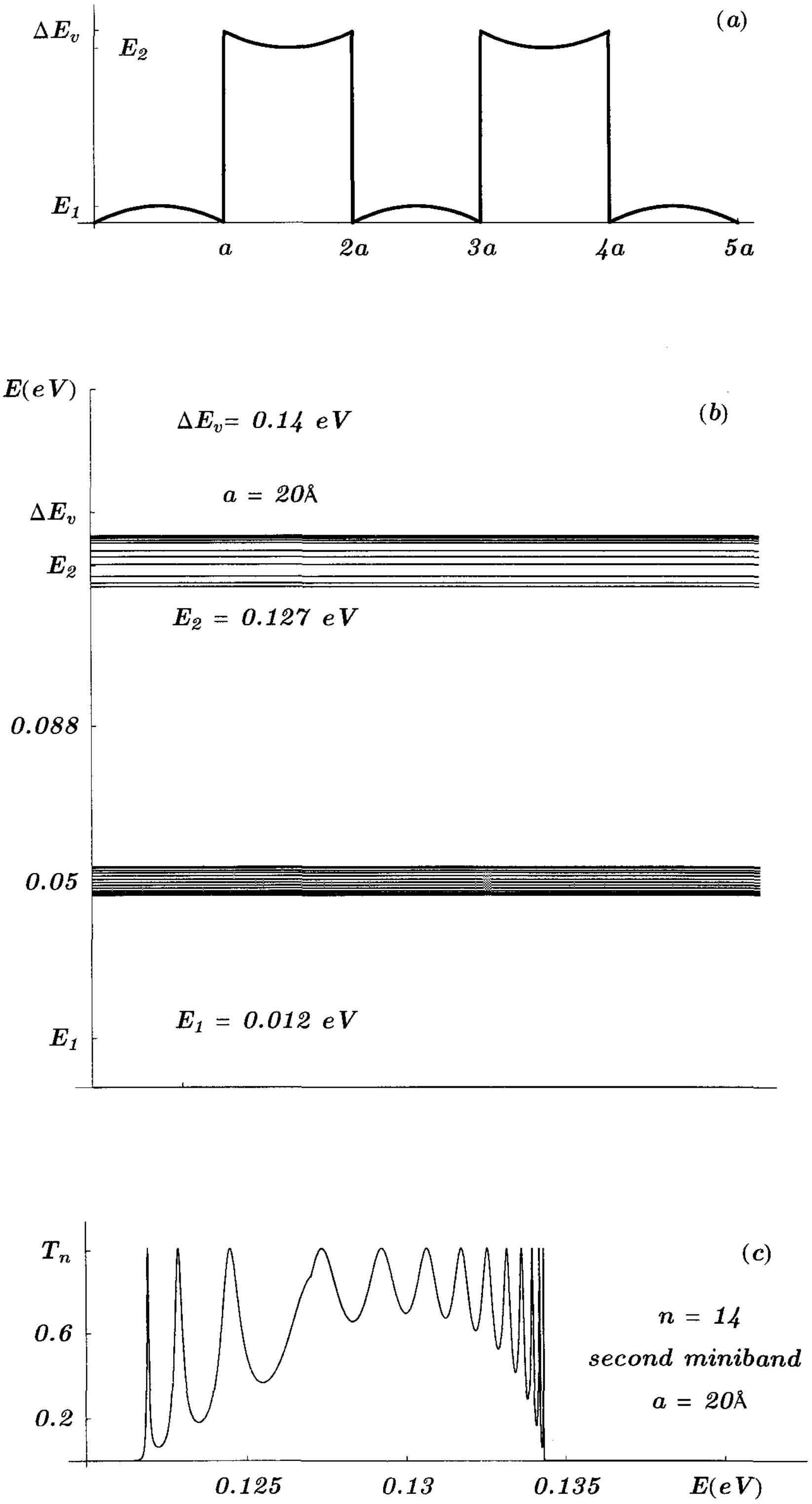

Fig.4

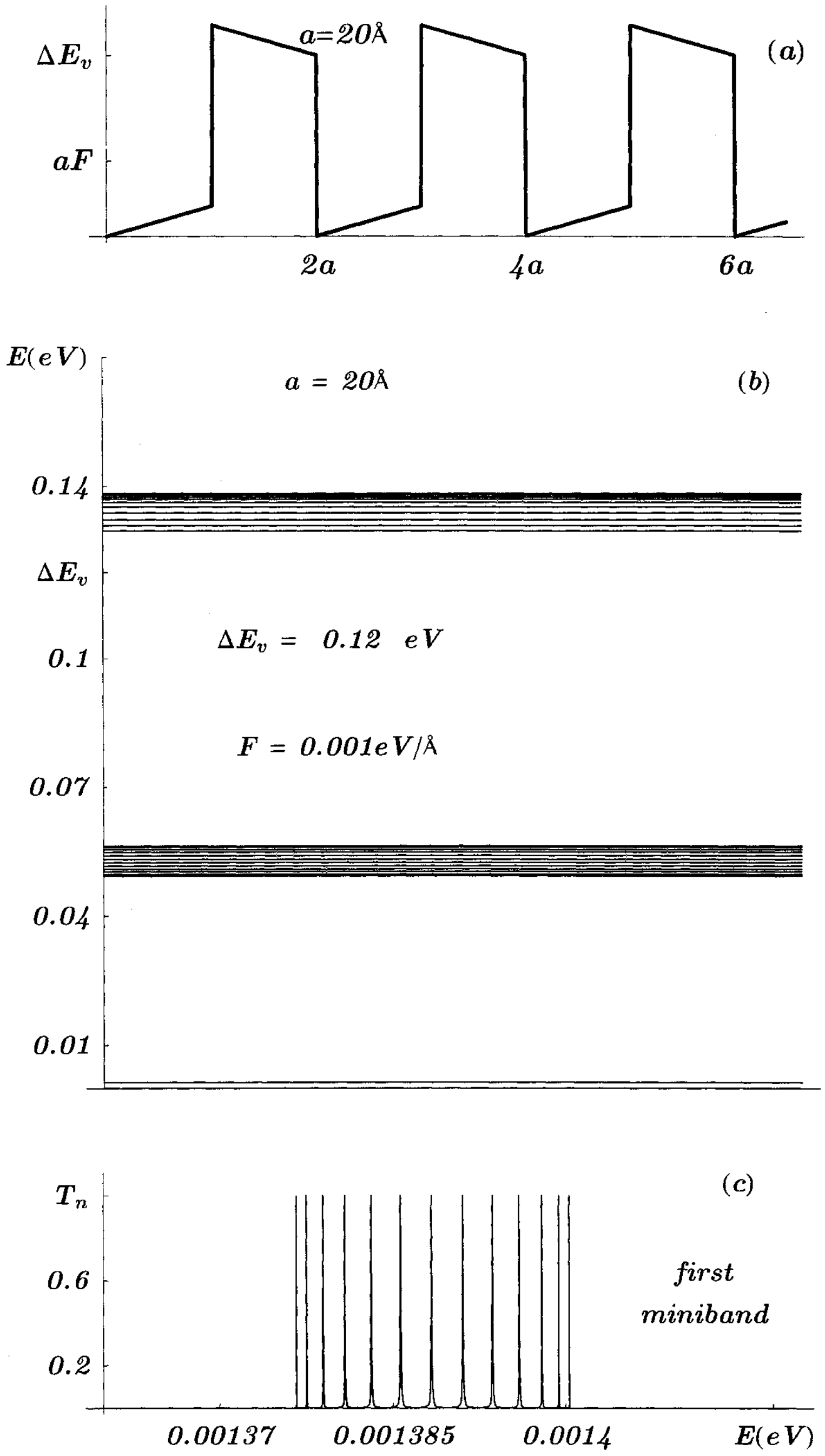

Fig.5

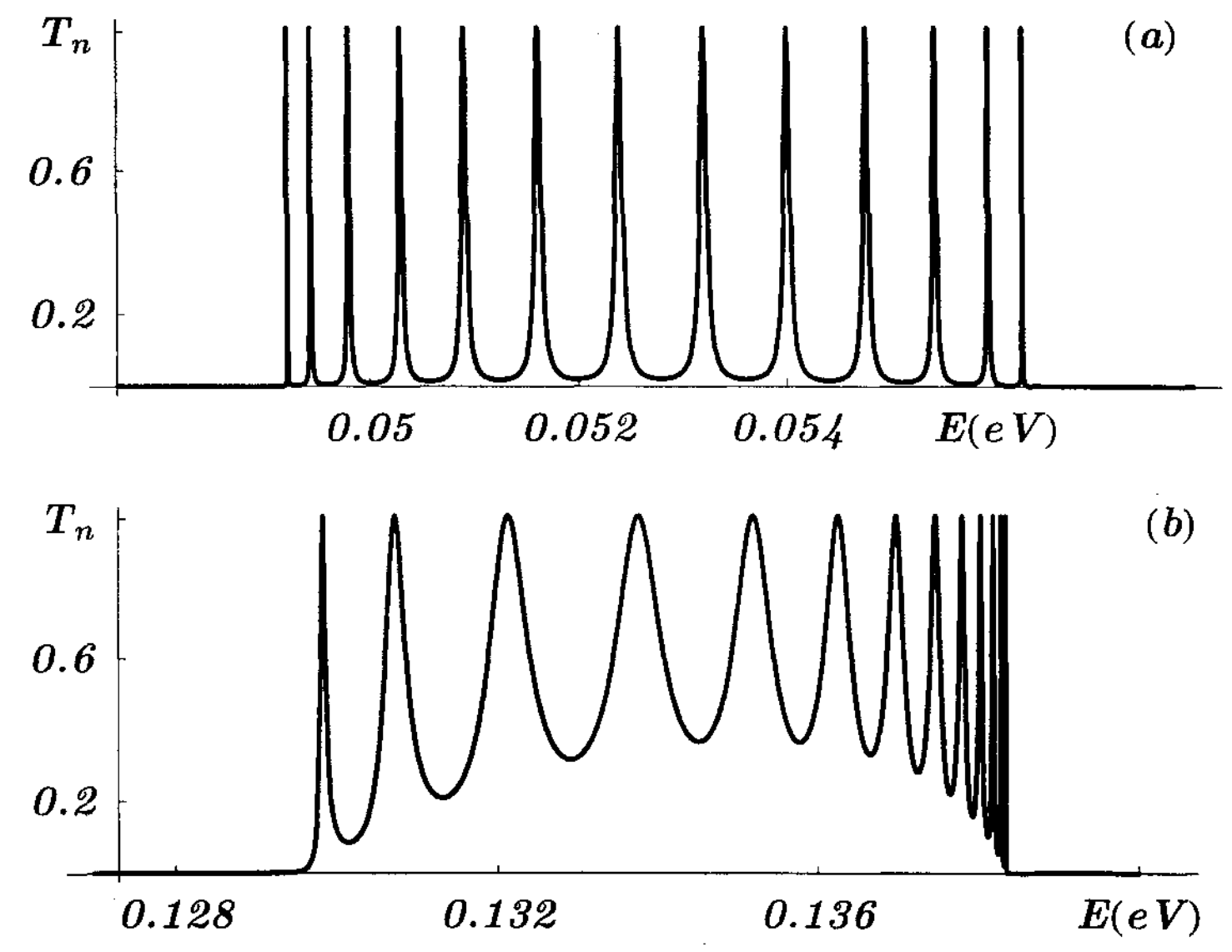

Fig.6

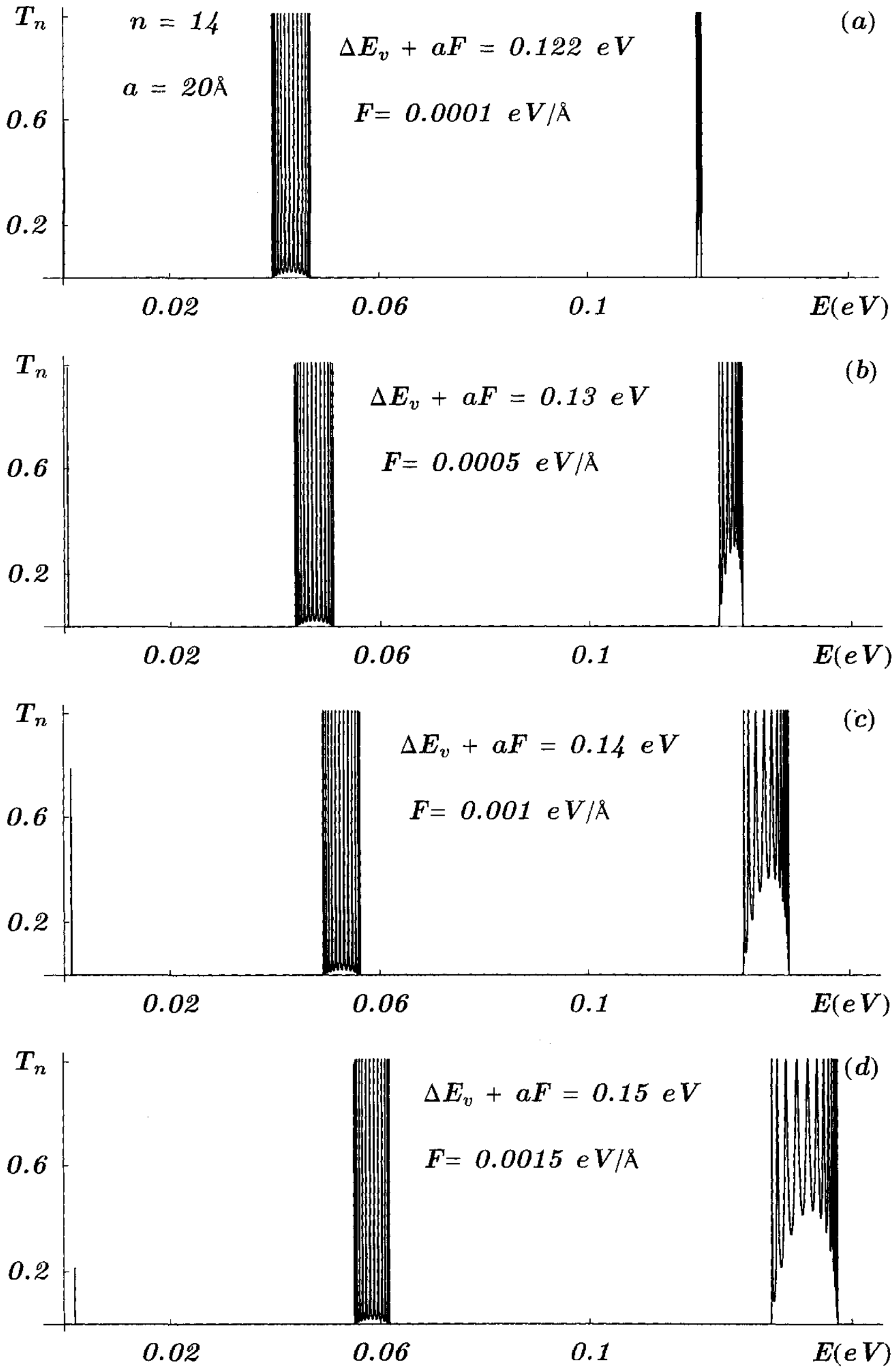

$T_n$
$n = 14$
$a = 20$Å
$\Delta E_v + aF = 0.122\ eV$
$F = 0.0001\ eV/$Å
(a)
0.6
0.2
0.02
0.06
0.1
$E(eV)$
$\Delta E_v + aF = 0.13\ eV$
$F = 0.0005\ eV/$Å
(b)
$\Delta E_v + aF = 0.14\ eV$
$F = 0.001\ eV/$Å
(c)
$\Delta E_v + aF = 0.15\ eV$
$F = 0.0015\ eV/$Å
(d)

Fig.7

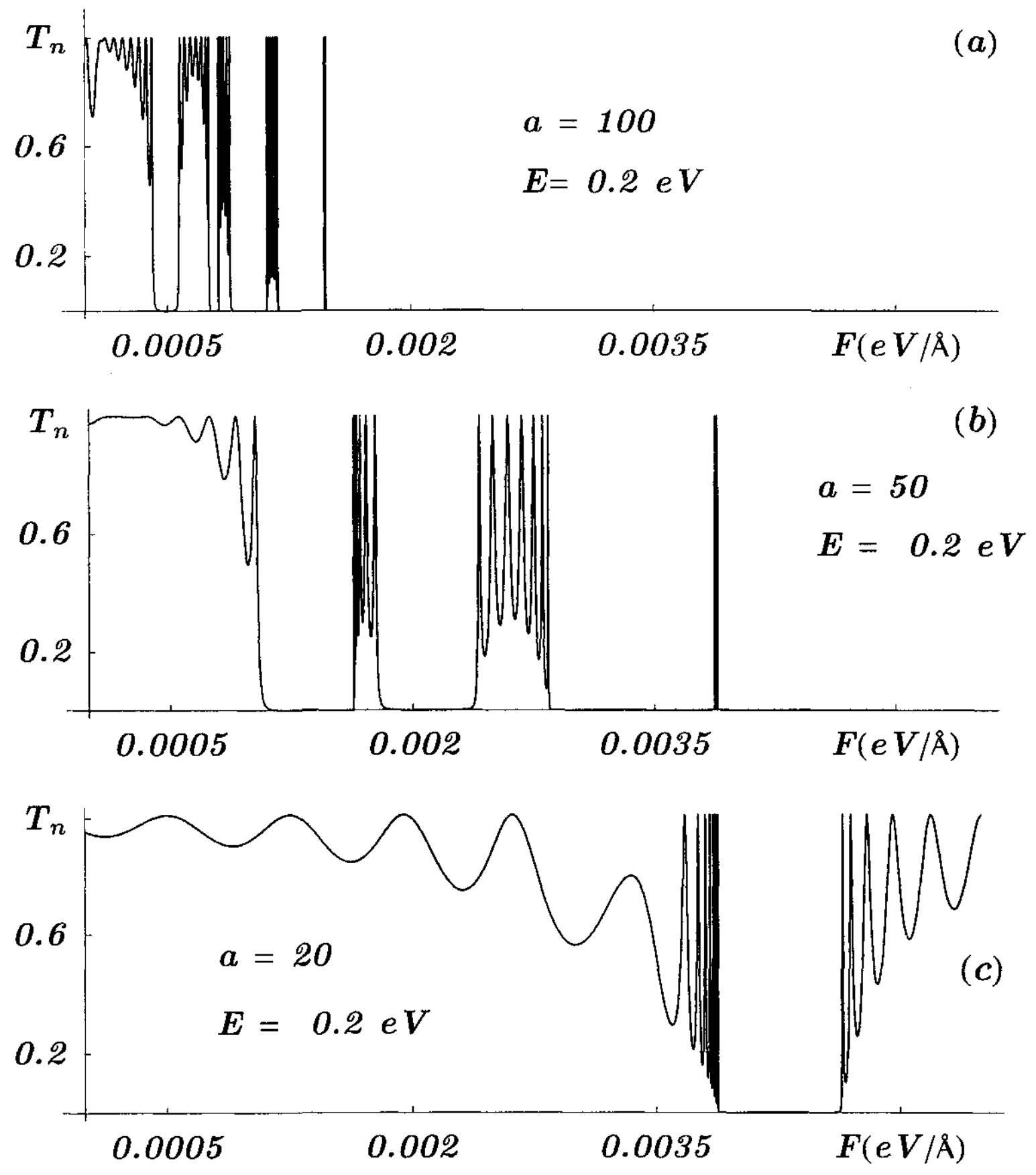

Fig.8

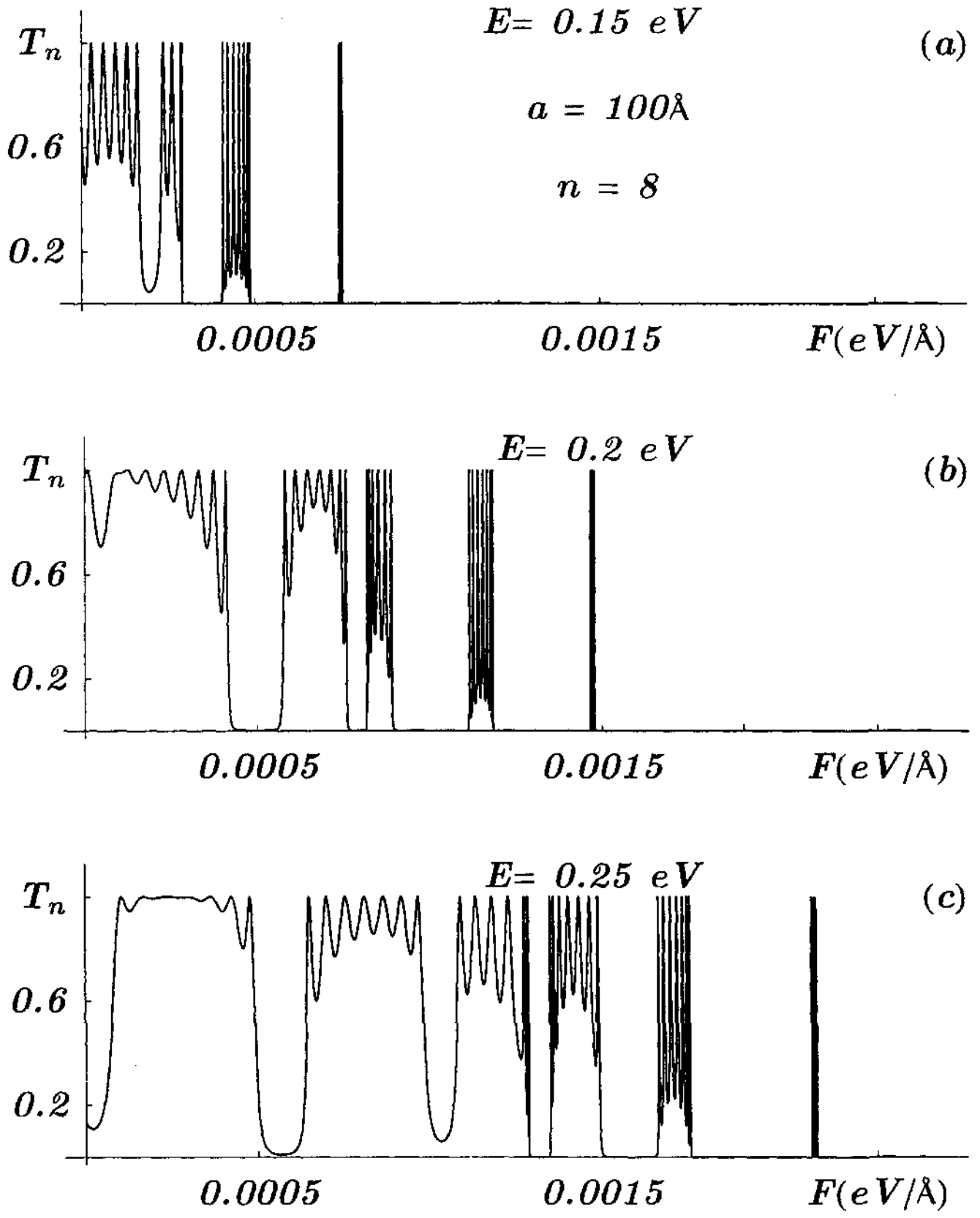

Fig.9

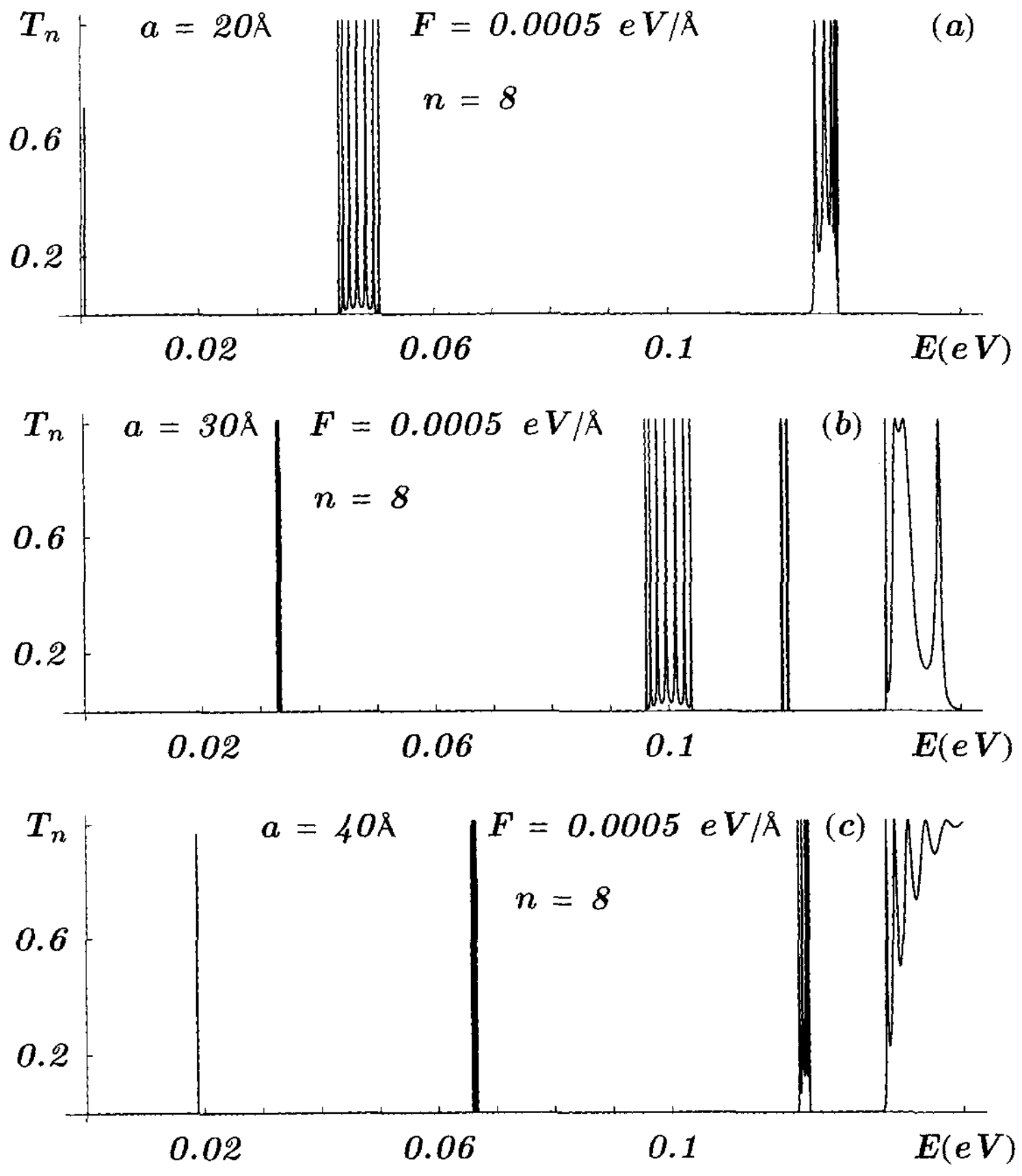

Fig.10

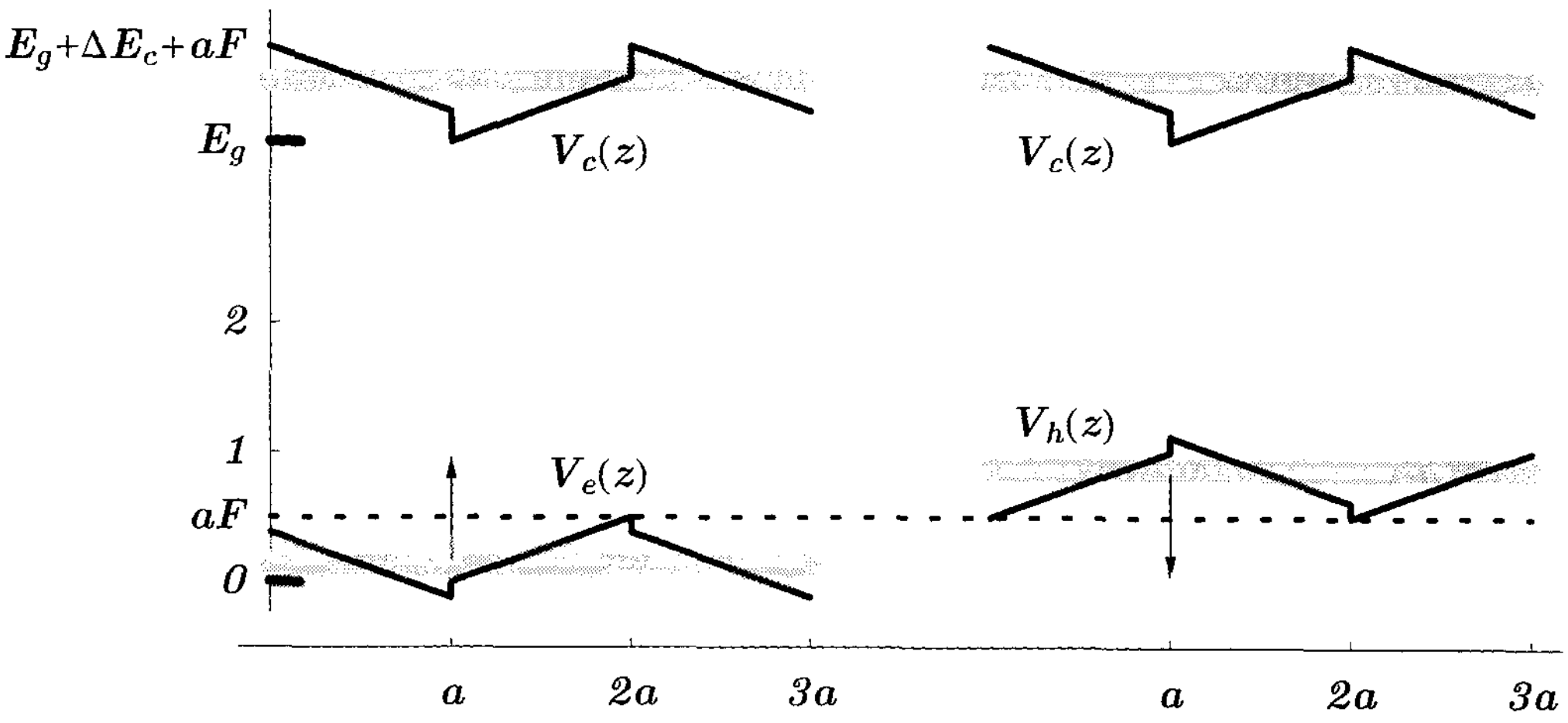

Fig.11

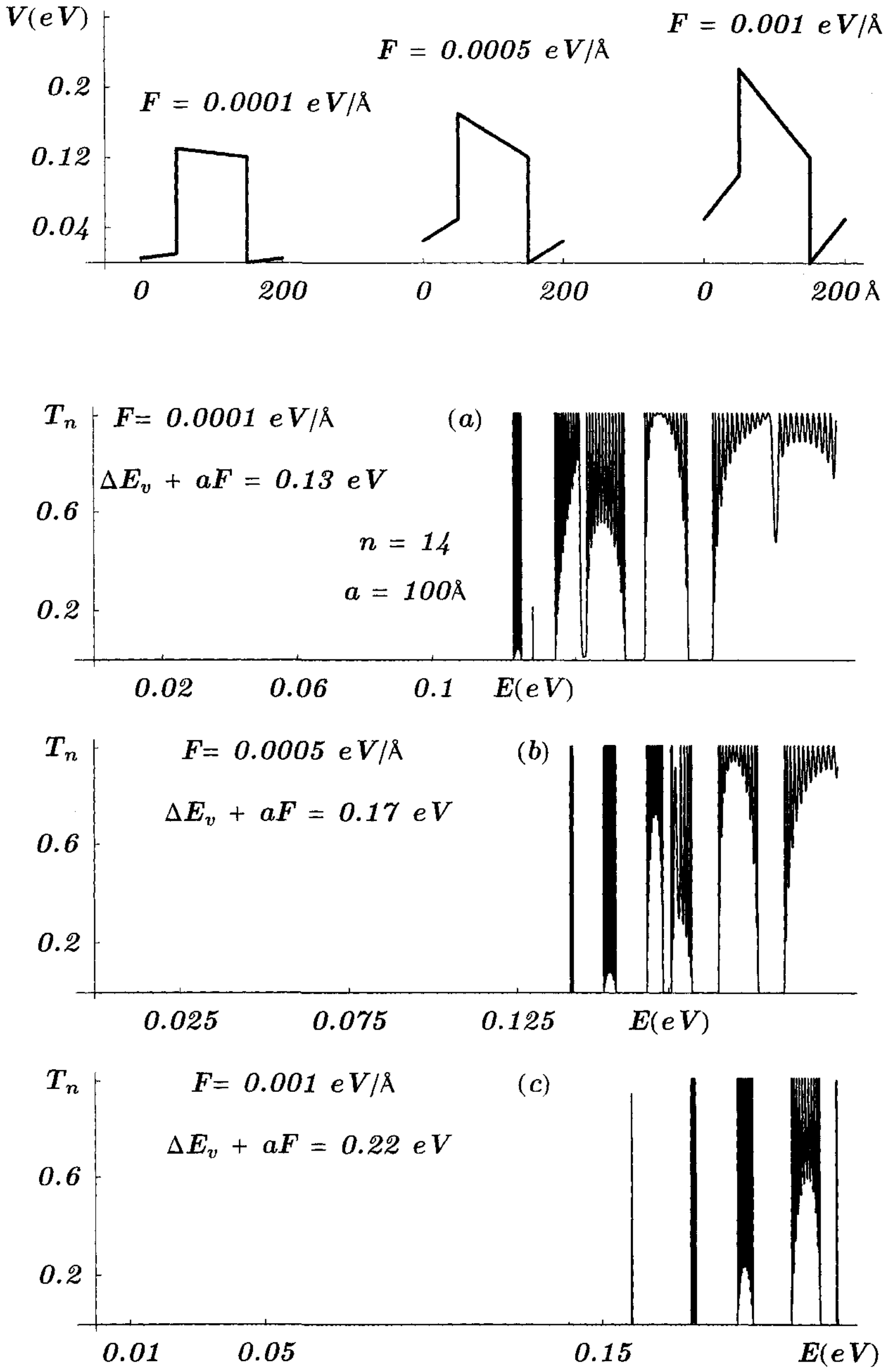

Fig.12